# Voiceprint recognition of Parkinson patients based on deep learning

Zhijing Xu[1], Juan Wang[1], Ying Zhang[1], Xiangjian He[2]

*[1]College of Information Engineering, Shanghai Maritime University, Shanghai, China*
zjxu@shmtu.edu.cn , wangjuan_y@foxmail.com , Jennif3r@foxmail.com
*[2]Global Big Data Technologies Centre, University of Technology Sydney, Sydney, Australia*
Xiangjian.He@uts.edu.au

## Abstract

*More than 90% of the Parkinson Disease (PD) patients suffer from vocal cord disorders. Speech impairment is an early indicator of PD. This study focuses on PD diagnosis through voiceprint features. In this paper, a method based on Deep Neural Network (DNN) recognition and classification combined with Mini-Batch Gradient Descent (MBGD) is proposed to distinguish PD patients from healthy people using voiceprint features. In order to extract the voiceprint features from patients, Weighted Mel Frequency Cepstrum Coefficients (WMFCC) is applied. The proposed method is tested on experimental data obtained by the voice recordings of three sustained vowels /a/, /o/ and /u/ from participants (48 PD and 20 healthy people). The results show that the proposed method achieves a high accuracy of the diagnosis of PD patients from healthy people, than the conventional methods like Support Vector Machine (SVM) and other mentioned in this paper. The accuracy achieved is 89.5%. WMFCC approach can solve the problem that the high-order cepstrum coefficients are small and the features component's representation ability to the audio is weak. MBGD reduce the computational loads of the loss function, and increase the training speed of the system. DNN classifier enhances the classification ability of voiceprint features. Therefore, the above approaches can provide a solid solution for the quick auxiliary diagnosis of PD in early stage.*

## 1. Introduction

Parkinson disease (PD) is the second common neurological disorder after Alzheimer disease [1]. Vocal characteristics are considered to be one of the earliest signs of this disease. At the early stage, the subtle anomalies of the sound are imperceptible to the listener, but the recorded speech signals can be acoustically analyzed for objective evaluation. The existing PD tests uses PET-CT imaging equipment to detect whether the dopaminergic neurons are reduced. But such tests are expensive and has high radiation levels, thereby reducing the accuracy of the diagnosis. Thus, a high accuracy, convenient, non-intrusive and inexpensive diagnostic method is required. In the 1990s, a variety of machine learning models were proposed, the most prominent one being Support Vector Machine (SVM) [2]. In 2015, Benba et al. proposed Mel Frequency Cepstrum Coefficient (MFCC) and SVM for voiceprint analysis of PD patients to distinguish between PD patients and healthy people [3]. In 2016, the authors [4] proposed to obtain voiceprints by extracting the cepstral coefficient of Relitive SpecTrAl Perceptual Linear Predictive (RSTA-PLP) and classify them with SVM classifier or *k*-Nearst Neighbor (*k*-NN) et al. Five different classifiers along with leave-one-subject-out scheme were used to distinguish between PD patients and patients with other neurological diseases. Later, they discussed the comparison of MFCC, PLP, and RSTA-PLP methods for extracting voiceprints and found that the highest accuracy was 90%, achieved by using PLP combined SVM classifier [5]. In addition, the authors [6] further studied the comparison of SVM's Multilayer Perceptron (MLP) kernel function with other kernel functions, the use of MFCC extraction features has very low-order cepstral coefficients, and the kernel function classifier has large computational complexity and long training time, and the accuracy of discrimination is 82.5% needing to be improved. In 2017, Benba et al. proposed a new improvement that used the Human Factor Cepstral Coefficient (HFCC) to extract the voiceprint features and the SVM classifier with linear kernel to obtain desired results [7]. In [8], Max A.Little used Linear Discriminant Analysis (LDA) to remove the remaining nuisance effects in the channel vectors and SVM or Probabilistic Linear Discriminant Analysis (PLDA) to classify the different types of distortion in Parkinson's voices. But SVM and other classifiers are nonparametric classifier with a shallow structure [9]. The drawback is that the ability to represent complex functions is limited in the case of finite samples and computational units [10], in comparison, there is a paper that uses deep learning and it has good accuracy of speaker recognition on a large scale of voiceprint corpus [11]. The recognition rate is stable and robust at about 95%, which means that it could be used in some real applications. Deep learning can achieve complex function approximation by learning a deep nonlinear network structure, and demonstrates the powerful ability to learn the essential features of data sets from a small sample set [12].

In recent years, with the advent of the big data era and the improvement of computer computing capability, deep learning has made breakthroughs in image recognition, machine translation, natural language processing and other applications. The core of deep learning is enabling the machine to imitate human activities such as audio-visual, thinking and learning. Therefore, it solves many complicated pattern recognition problems and makes great progresses in correlation technique of Artificial Intelligence (AI). As a widely used Artificial Neural Network (ANN) model in deep learning, Deep Neural Network (DNN) has made lots of achievements in issues such as signal processing, language modeling and bioinformatics [13,14,15]. Its nonlinear structure has powerful modeling capability.

According to the ideas and advantages or disadvantages of various algorithms discussed above, in this paper we propose a new DNN-based voiceprint recognition model in this paper, which can use voiceprint features to discriminate between PD patients and healthy people and detect the disease in its early stages.

The main contributions of this paper are:

● Weighted-MFCC (WMFCC) is used to extract the voiceprint features to enhance the sensitive components, and then obtain the parameters of PD patients' dysphonia detection. The WMFCC approach, based on the entropy method, solves the problems that the high-order cepstrum coefficients are small and the feature component's representation ability to the audio is weak.

● Multi-layer neural network recognition and classification of the DNN in deep learning are used to improve the accuracy of discriminating PD patients. When the DNN is training, the model obtains better initial parameters through unsupervised parameter pre-training algorithm. On this basis, the model uses the supervised training method to optimize parameters further.

● Model optimization based on Mini-Batch Gradient Descent (MBGD) optimization algorithm is proposed to reduce the amount of computation of the loss function, and increase the training speed of the system.

● Compared with the approaches based on the traditional Support Vector Machine (SVM) and other state-of-art classification methods for testing and classifying the samples in the PD dataset, it is found that our approach achieves the highest accuracy in diagnosing in the Parkinson disease.

In the next section, we will introduce the dataset and the technique of acquisition. Section 3 will describe our proposed approach for feature extraction, model optimization, recognition and classification. In Section 4, our experiments and results will be presented. Section 5 will provide a discussion in which the results will be interpreted in depth with a highlight on the comparison of our research work with the state-of-art work. The conclusion is presented in Section 6.

## 2. Datasets

The dataset used in this study was collected and used in [16]. It contains 20 (6 women and 14 men) patients with PD and 20 (10 women and 10 men) healthy people. For PD patients the time since diagnosis ranges between 0 and 6 years, and the age of patients ranges between 43 and 77 (mean: 64.86, standard deviation: 8.97). The age of healthy people ranges between 45 and 83 (mean: 62.55, standard deviation: 10.79). All the recordings were performed by a Trust MC-1500 microphone with a frequency range between 50 Hz and 13 kHz. The microphone was set to 96 kHz, 30 dB and placed at a 15 cm distance from subjects. All the samples were made in stereo-channel mode and saved in WAV format. In this study, we use 3 types of recordings, which were obtained by inviting each person of the 40 participants (20 PD and 20 healthy) to pronounce three sustained vowels /a/, /o/ and /u/ at a comfortable level. This gives us a dataset containing 120 voice samples. The analyses are made on these samples.

An independent dataset was also collected using the same recording devices with the same physicians. 28 PD patients were invited to pronounce sustained vowels /a/ and /o/ three times, and then we selected only one sample of the two vowels. For these patients, the time since diagnosis ranges between 0 and 13 years, and the age of patients ranges between 39 and 79 (mean: 62.67, standard deviation: 10.96). This dataset was used to test and validate the obtained results using the first dataset.

## 3. Methodology

Using the voiceprint features of PD patients to recognize and classify healthy people, this paper builds a DNN-based PD patient recognition model, as shown in Fig. 1.

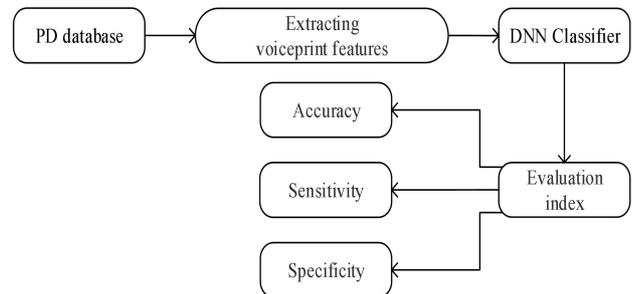

Fig. 1. PD patients identification system model

### 3.1. WMFCC voiceprint feature extraction

The extraction of speech feature parameters is crucial in voiceprint recognition. In the field of voiceprint recognition, the most commonly used feature for extraction is MFCC [17]. The speech signal changes at a slower rate. When it is perceived in a short time, the speech signal is generally considered to be stable at intervals of 10-30 ms [18]. Therefore, it should be calculated by short-time spectrum analysis. The Mel scale is used to estimate the frequency perception of the human ear, and it is calculated by 1000 Hz corresponding to 1000 Mel.

This paper uses temporal speech quality, spectrum, and cepstrum domains to develop more objective assessments to detect speech impairments [19]. These measurements include the fundamental frequency of vocal cord vibration ($F_0$), absolute sound pressure level, jitter, shimmer and harmonics noise ratio (HNR). From [20], the details shown in table 1.

Table 1. Acoustic analysis results of healthy male and female, male and female with PD

| Groups | Gender | Average age | $F_0$(Hz) | Jitter (%) | Shimmer (%) | HNR (dB) |
|---|---|---|---|---|---|---|
| Healthy | male | 58.4 ± 12.5 | 128.4 ± 17.6 | 0.04 ± 0.36 | 0.26 ± 0.10 | 14.8 ± 4.6 |
| Healthy | female | 55.6 ± 11.9 | 205.4 ± 37.6 | 1.16 ± 1.15 | 0.35 ± 0.46 | 11.0 ± 7.1 |

| PD | male | 61.2 ± 9.6 | 120.5 ± 20.8 | 0.94 ± 0.76 | 0.37 ± 0.16 | 10.4 ± 3.7 |
| PD | female | 61.7 ± 10.6 | 193.8 ± 16.4 | 1.94 ± 1.30 | 0.68 ± 0.91 | 8.1 ± 5.1 |

Based on the pronunciation characteristics of PD patients, the characteristic parameters were extracted for analysis. However, each component contained in the feature parameters has different voiceprint characterization capabilities for different speech samples. The traditional MFCC method extracts the voiceprint features with low-order cepstral coefficients and the feature components have poor representational capabilities for audio. In order to enhance the sensitive components of recognition, this paper analyzes the contribution of each dimension feature parameter to the voiceprint representation by calculating the entropy value of multi-dimensional corpus, and extracts the voiceprint features by the entropy weighted method, thus improving the recognition accuracy of the system. Fig. 2 is a flow chart of the voiceprint feature extraction of WMFCC.

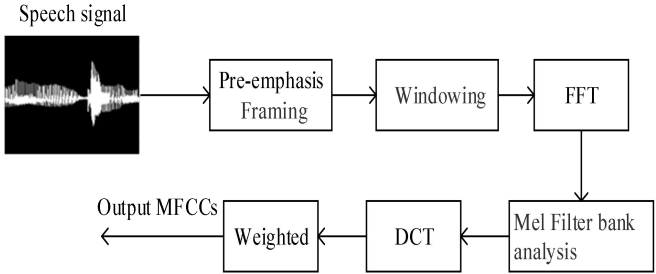

Fig. 2. Flow chart of WMFCC extraction

The detailed extraction process is as follows

### 3.1.1. Pre-emphasis, framing

In order to eliminate the effect of lips and vocal cords during the vocalization process, compensating the high-frequency part of speech signal suppressed by the vocal system and highlighting the high-frequency formant. Therefore, the first-order differential equation is applied to the speech samples to increase the amplitude of the high frequency formant [21]. In fact, the speech signal is passed through a high-pass filter:

$$H(z) = 1 - kz^{-1}, \qquad (1)$$

where $k$ is the pre-emphasis coefficient and it should be in the range of [0,1] (usually 0.97).

In the framing step, the speech signal is divided into $N$ sample frames. In order to avoid the excessive change of two adjacent frames, there is an overlapping area, containing $M$ sampling points with $M < N$, between two adjacent frames.

### 3.1.2. Hamming window

The purpose of adding a Hamming window is to reduce signal discontinuities and make the end smooth enough to connect with the beginning.

Assume that the signal after the framing is $s_n$, where $n$ is the size of the frame, of which $\{s_n, n = 1, ..., N\}$. The form of $s_n^{'}$ is as:

$$s_n^{'} = \left\{ 0.54 - 0.46 \cdot \cos\left(\frac{2\pi(n-1)}{N-1}\right)\right\} \cdot s_n. \quad (2)$$

### 3.1.3. Fast Fourier transform

The Fast Fourier Transform (FFT) is used to convert $N$ samples from the time domain to the frequency domain. The FFT is used because it is a fast algorithm that implements the Discrete Fourier Transform (DFT). The DFT is defined on $N$ sample sets, and the DFT of the speech signal is:

$$S_n = \sum_{k=0}^{N-1} s_k e^{-j2\pi kn/N}, \, n = 0,1,2,..., \, N-1, \quad (3)$$

where $s_k$ is an input speech signal and $N$ is the number of sampling points of the Fourier Transform.

### 3.1.4. Filter bank analysis

There are several redundant signals in the frequency domain, and the filter bank can streamline the amplitude of the frequency domain. The human ear's perception of sound is not linear. It is better described by the nonlinear relationship of log. The relationship between Mel frequency and speech signal is as shown in Equation (4) [22].

$$Mel(f) = 2595\ln(1 + \frac{f}{700}), \qquad (4)$$

where $Mel(f)$ represents the Mel frequency, whose unit is mel, and $f$ is the frequency of the speech signal, whose unit is Hz.

### 3.1.5. Logarithm/Discrete Cosine Transform

In this phase, the MFCC is calculated from the log filter bank amplitudes $(m_j)$ through the Discrete Cosine Transform (DCT) [7]:

$$c_i = \sqrt{\frac{2}{N}} \sum_{j=1}^{N} m_j \cdot \cos\left(\frac{\pi i}{N}(j - 0.5)\right), \qquad (5)$$

where $N$ is the number of filter bank channels, $m_j$ is the amplitude of the $j$-th log fliter bank.

### 3.1.6. Weighted

The main advantage of cepstral coefficients is they are not related to each other. Thus it is convenient to analyze the cepstral coefficients of each order. As the high-order cepstral coefficients are very small, and the sensitive components are not obvious, they reduce the recognition rate of the extracted effective features and the subsequent classification recognition rate, as shown in Fig. 3. Therefore, based on the MFCC method, the entropy method is used to improve the characterizing ability of the feature components to the voiceprint features. This method is simple and considers the interaction among the feature components.

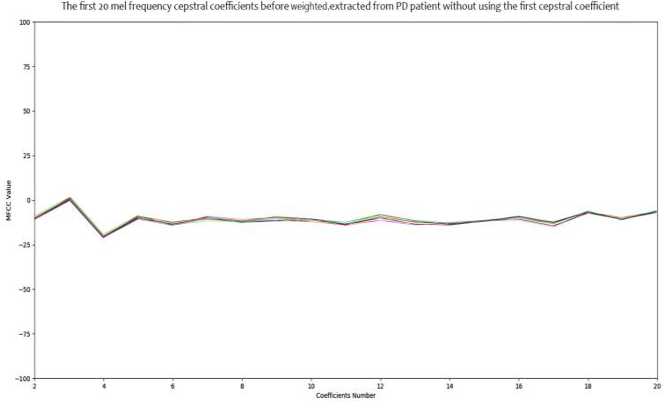

Fig. 3. The first 20 mel frequency cepstral coefficients of a PD patient before weighted (without using the first cepstral coefficient)

The entropy method is an objective weighting method for calculating the weights among mutually independent variables. The weight of the components is determined according to the information entropy of the calculated components [23]. The larger the value of entropy, the less information is carried and the smaller the weight of the component is. On the contrary, the smaller the value of entropy, the more information is carried and the larger the weight of the component is. Therefore, it is a crucial step in changing these cepstral coefficients (Fig. 4). It is achieved in the following.

In the voiceprint features of the PD database speech sample

$$MFCC = (M_1, M_2, M_3, ..., M_i, ..., M_N),$$  (6)

where $M_i = (mel_{i(1)}, ..., mel_{i(j)}, ..., mel_{i(D)})$ is the feature vector of the $i$-th frame of the voiceprint feature, $D$ is the feature parameter dimension, $N$ is the number of frames of the speech sample, and $mel_{i(j)}$ is the $j$-th feature vector value of the $i$-th frame of the voiceprint feature.

First, the feature matrix is normalized as shown in Equation (7):

$$mel_{i(j)}^{''} = \frac{\max\{mel_j\} - mel_{i(j)}}{\max\{mel_j\} - \min\{mel_j\}}$$  (7)

The definition of entropy is as shown in Equation (8),

$$e_j = -k \sum_{i=1}^{N} Y_{ij} \times \ln Y_{ij},$$  (8)

where $Y_{ij} = \dfrac{mel_{i(j)}^{''}}{\sum\limits_{i=1}^{N} mel_{i(j)}^{''}}$,

Equation (9) is the entropy weight of the obtained feature component:

$$w_j = \frac{1 - e_j}{\sum\limits_{j=1}^{D} (1 - e_j)}$$  (9)

Finally, the weights of the components of the MFCC are calculated by Equation (9), and the new parameters obtained are as shown in Equation (10):

$$wM_i = (w_1 \cdot mel_{i(1)}, ..., w_D \cdot mel_{i(D)})$$  (10)

Taking a speech sample as an example, the feature corresponding to the cepstral coefficients of the first 20 mel frequency of the PD patient are extracted, and the weight of the feature components are calculated by the entropy weighting method, as shown in Fig. 4.

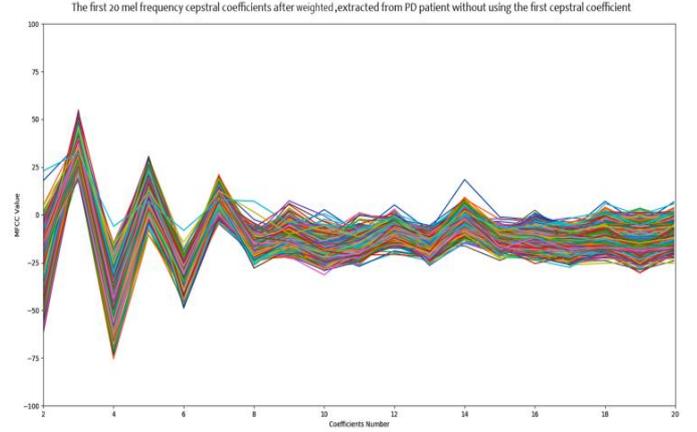

Fig. 4. The first 20 mel frequency cepstral coefficients of a PD patient after weighted (without using the first cepstral coefficient)

The multi-cepstral coefficients of the WMFCC are extracted in each of the obtained speech samples, and the extracted coefficients range from 1~20. Note that the first cepstral coefficient loses the reference meaning due to large amplitude change. We continue to obtain the optimum value of the coefficient required for the best classification accuracy in this way. Next, the corresponding voiceprint is extracted by calculating the average of all the frames to obtain each voiceprint, as shown in Fig. 5.

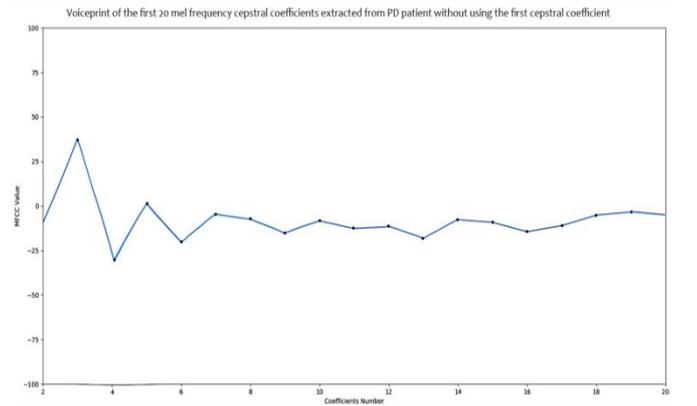

Fig. 5. Voiceprint of the first 20 mel frequency cepstral coefficients of a PD patient (without using the first cepstral coefficient)

Comparing Fig. 3 and Fig. 5, it can be intuitively concluded that WMFCC solves the problem that the high-order cepstral coefficients are very small. After the weighted average, the sensitivity of the MFCC parameters is also highlighted, and the change of the high-order cepstral coefficients affects the recognition rate of the subsequent effective features.

### 3.2. DNN Classification

According to the voiceprint features in the dataset extracted by the WMFCC, the DNN training method is used for feature classification. The model structure, training process and the network optimization algorithm of DNN described as follows.

### 3.2.1. DNN structure

DNN is a multilayer perceptron with multiple hidden layers. Because it contains multiple hidden layers, it can abstract useful high-level features or attributes from high-redundant low-level features, and then discover the inherent distribution of data [24]. The neural network designed in this paper includes input layer, hidden layer and output layer. As shown in Fig. 6, the input layer is written as Layer 0, while the output layer is written as layer L.

A DNN can have multiple hidden layers, and the output of the current hidden layer is the input of the next hidden layer or the output layer. We use the Back-Propagation (BP) algorithm to calculate the gradient of each layer's parameters. The activation function is a Rectified Linear Unit (ReLU), which has the advantage that the network can introduce sparsity on its own and greatly improve the training speed.

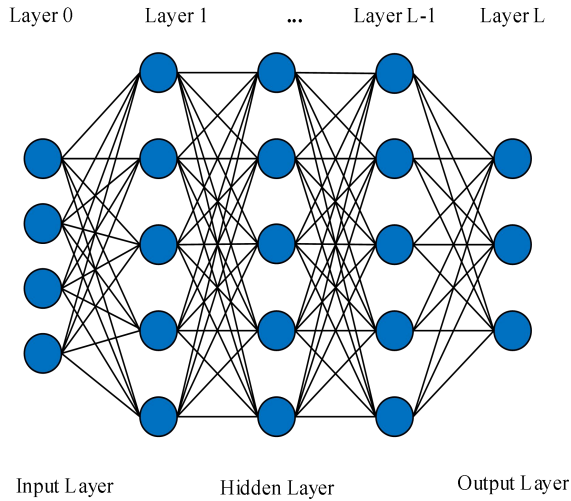

Fig. 6 Schematic diagram of the DNN structure

For any $l\left(0 < l < L\right)$ layer,

$$z^l = W^l v^{l-1} + b^l,\qquad(11)$$

$$v^l = f\left(z^l\right),\qquad(12)$$

where $z^l \in R^{N_l \times 1}$ is the excitation vector, $v^l \in R^{N_l \times 1}$ is the activation vector, $W^l \in R^{N_l \times N_{l-1}}$ is the weight, $b^l \in R^{N_l \times 1}$ is the bias, and $N_l \in R$ is the number of neurons in the $l$ th layer. $f(\cdot)$ is the activation function ReLU, with a mathematical expression:

$$\mathrm{Re}LU(z) = \max(0, z).\qquad(13)$$

### 3.2.2. Parameter Pre-training Algorithm

The DNN pre-training uses a layer-by-layer pre-training method based on a Restricted Boltzmann Machine (RBM). The DNN is treated as a Deep Belief Network (DBN) with a number of RBMs stacked, and then pre-trained bottom-up layer by layer [25]. The detailed process is as follows.

If the input is a continuous feature, a Gauss-Bernoulli-distributed RBM is trained, and if the input is a binomial distribution feature, a Bernoulli-Bernoulli-distributed RBM is trained. The output of the hidden layer is then used as input data for the next layer of either the Gauss-Bernoulli-distributed RBM or Bernoulli-Bernoulli distribution RBM, depending on the input feature. This process does not require label information and is an unsupervised training process. Supervised training is performed after pre-training. According to the task and application requirements of this study, the label of the training data and the output of the criterion are added at the top level, and the back propagation algorithm is used to adjust the parameters of the network.

### 3.2.3. Back Propagation Algorithm

When using back propagation for parameter training, the model parameters of the DNN are trained through a set of training $\left(x^i, y^i\right), 1 \leq i \leq N$, where $x^i$ is the feature vector of the first $i$ samples, and $y^i$ is the corresponding label. The back propagation algorithm is explicitly summarized below.

1. Input $x$ : Set the corresponding activation value for the input layer.

2. Forward Propagation: Calculate the corresponding Equations (11) and (12) for each layer.

3. Output layer $e^L$ : The error vector is calculated by:

$$e^L = \frac{\partial J\left(W, b; x, y\right)}{\partial z^L}\qquad(14)$$

4. Back propagation: The error of defining the layer 1 node is:

$$e^l = diag\left(f_l'\left(z^l\right)\right) \cdot \left(W^{l+1}\right)^T \cdot e^{l+1}\qquad(15)$$

5. Output: The weight matrix and bias of each layer are calculated by Equations (16) and (17), respectively.

$$x(n) = x(0)\hat{x}(n) + \sum_{k=0}^{n-1}\frac{k}{n}x(k)x(n-k)\qquad(16)$$

$$\begin{cases} 0 & ,\ n<0 \\ \dfrac{x(n)}{x(0)} - \sum_{k=0}^{n-1}\dfrac{k}{n}\hat{x}(k)\dfrac{x(n-k)}{x(0)} & ,\ n>0 \end{cases}\qquad(17)$$

### 3.2.4. Small batch gradient descent optimization algorithm

The BP algorithm is the core algorithm for training DNN. It optimizes the parameter values in the network according to the predefined loss function. An important step in determining the quality of the network model is the optimization of the parameters in the neural network model [26].

Among them, the Gradient Descent (GD) solves the minimum value of the loss function and solves it step by step through the algorithm to obtain the minimum loss function and model parameter values. Since each update of the parameters will traverse all sample data, it is a complete gradient drop and has a high accuracy. However, it will spend a lot of time on traversing the collection.

However, the Stochastic Gradient Descent (SGD) algorithm does not optimize the loss function on all training samples, but randomly optimizes the loss function of a training sample at each iteration [27]. As a result, the speed of each round of parameter updates will be greatly accelerated. However, since the SGD algorithm optimizes the loss function on a certain sample each time, the disadvantage is obvious, a small loss function of the local sample does not mean that the loss function of all samples is small. The neural network obtained by stochastic gradient descent optimization is difficult to achieve global optimization, and the accuracy is lower than the complete gradient descent algorithm.

In order to solve the shortcomings of the gradient descent algorithm and the stochastic gradient descent algorithm, in this paper, a new fusion algorithm, Mini-Batch Gradient Descent (MBGD) algorithm, is proposed, and it only calculates the loss function of a small number of training samples when updating each parameter. A small portion of the sample is referred as a batch in this paper. On one hand, using matrix operations, optimizing the parameters of a neural network on a batch is comparable to a single sample. On the other hand, using a small portion of the sample each time, the number of iterations required for convergence can be greatly reduced. When the convergence is reduced, the accuracy obtained is closer to the result of the gradient descent algorithm. Therefore, the MBGD algorithm can overcome the shortcomings of the above two algorithms, and at the same time take into account their advantages.

The MBGD algorithm randomly extracts $m$ samples from all samples, where $m$ is the total number of training samples. The $m$ samples are $X_1, X_2, ..., X_i, ..., X_m$. $\omega$ and $b$ are the sets of weights and bias in the network. $Y_i$ and $A_i$ are the expected output and the actual output of the first $i$ samples input, and $\|\bullet\|$ is a norm operation. The mean squared error is calculated as follows:

$$C(\omega, b) = \frac{1}{2m} \sum_{i=1}^{m} \|Y_i - A_i\|^2 = \frac{1}{m} \sum_{i=1}^{m} C_{X_i}. \quad (18)$$

Among $C_{X_i} = \dfrac{\|Y_i - A_i\|^2}{2}$. According to the gradient, the representation of $\nabla C$ is:

$$\nabla C = \frac{1}{m} \sum_{i=1}^{m} \nabla C_{X_i} \quad (19)$$

Equation (19) estimates the overall gradient using $m$ sample data, and larger the $m$ is, the more accurate the estimates' result is. At this point, the formula for the update are:

$$\omega_k^{'} = \omega_k - \eta \frac{\partial C}{\partial \omega_k} = \omega_k - \frac{\eta}{m} \sum_{i=1}^{m} \frac{\partial C_{X_i}}{\partial \omega_k}, \quad (20)$$

$$b_k^{'} = b_l - \eta \frac{\partial C}{\partial b_l} = b_l - \frac{\eta}{m} \sum_{i=1}^{m} \frac{\partial C_{X_i}}{\partial b_l}. \quad (21)$$

where $\eta$ is a positive number, whose range of value is [0,1], and is called the learning rate.

Based on the MBGD optimization algorithm, a flow chart of the MBGD optimization algorithm is drawn, as shown in Figure 7.

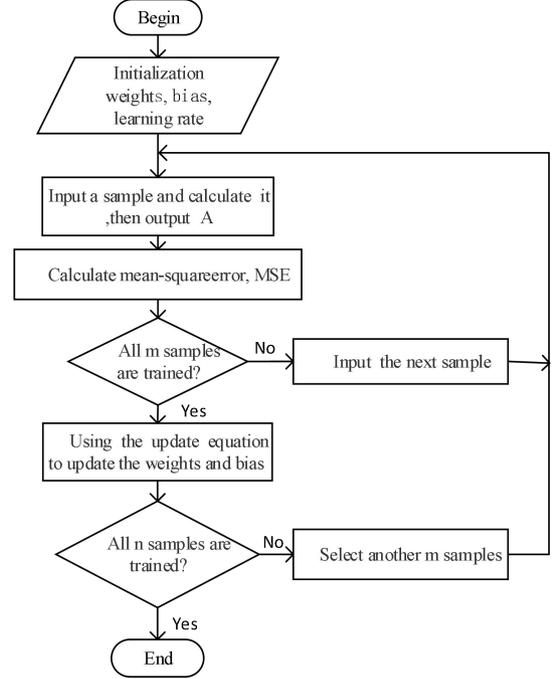

Figure 7 Flow chart of the MBGD optimization algorithm

The PD database is known to have a total of 120 samples. Since the total number of samples is limited, after several trials, it is finally determined that two samples are taken each time as a batch to calculate the loss function and update the parameters. After 60 iterations, the training of the entire speech sample set is completed, and this is called an epoch. Since the loss function is calculated by using multiple samples for each update, the calculation of the loss function and the updates of the parameters are more informative, the decline of the loss function is more stable, and the convergence speed is faster. At the same time, the use of small batch calculations also reduces the amount of calculation. This process is calculated by Equations (18-21).

## 4. Experimental results

This paper uses the compressed frames of 20 healthy people and 20 patients with PD to train, using $k$-fold cross validation. This method is used to measure the predictive performance of the built model. The training model has excellent performance on the new data. On one hand, it can greatly reduce the over-fitting. On the other hand, it can obtain as much more valid information from the limited sample data as possible. This method is a cross-validation method that is used when the sample size is small. The $k$-fold cross-validation works as follows.

The initial sampling is divided into $k$ sub-samples, of which a single sub-sample is taken as the data of the verification model, the remaining $k$-1 samples are used for the training. This process is repeated $k$ times, while each sub-sample is verified once. After that, the obtained $k$ results are averaged to evaluate the performance of the model. When $k$=$n$ (where $n$ is the amount of total samples), it is called the leave-one method [3,5]. The test set for each training requires only a single sample, and a total of $n$ times trainings and predictions are performed. The training samples selected using this method are only one sample fewer than the total dataset, so they are closest to the distribution of the original samples.

The proposed approach is also tested using an independent test set of 28 PD patients collected by the same physician. Furthermore, it is compared with the SVM method of different kernels studied by Benba et al., namely, Radial Basis Function (RBF), Linear, Polynomial (POL) and MLP on SVM classifier [6]. To test the success rate of these classifiers in identifying PD patients and healthy people, the accuracy, sensitivity and specificity are calculated. Accuracy represents the success rate of distinguishing between two groups of participants, sensitivity represents the accuracy of detecting healthy people, and specificity represents the accuracy of detecting PD patients. TP is true positive (healthy people are classified correctly), TN is true negative (PD patients are classified correctly), FP is false positive (PD patients are classified incorrectly), and FN is false negative (healthy people are classified incorrectly). In addition, there are two evaluation criteria Matthews Correlation Coefficient (MCC) and Probability Excess (PE), needed to be further calculated to show the quality of binary classification. The calculation formula are Equations (22-26)

$$Accuracy = \frac{TP + TN}{TP + TN + FP + FN} \quad (22)$$

$$Sensitivity = \frac{TP}{TP + FN} \quad (23)$$

$$Specificity = \frac{TN}{TN + FP} \quad (24)$$

$$MCC = \frac{TP \cdot TN - FN \cdot FP}{\sqrt{((FN+TP)(FP+TN)(FP+TP)(FN+TN))}} \quad (25)$$

$$PE = \frac{TP \cdot TN - FN \cdot FP}{(FN + TP)(FP + TN)} \quad (26)$$

Preprocessing and feature extraction are performed on each type of speech samples of /a/, /o/ and /u/ . Then, the DNN is used to train and calculate evaluation indicators such as accuracy, sensitivity, specificity, MCC and PE.

Table 2 shows the classification accuracy of the vowels /a/ in the same data set for each classifier. In the DNN classification method, the characteristics of the extracted 12-th and 14-th Mel frequency cepstral coefficients achieve the maximum classification accuracy of 84.5%. Therefore, 33 people are correctly classified, and only 7 are misclassified. Using the same parameters, the maximum sensitivity is 85% and the specificity is 80%. It is concluded that 17 healthy people and 16 PD patients are classified correctly. The maximum MCC is 0.6917 and the PE is 0.6500.

Table 2. Results using vowel /a/

| Classi-fier | Accur-acy(%) | Sensiti-vity(%) | Specifi-city(%) | MCC | PE | Coeffi-cients |
|---|---|---|---|---|---|---|
| RBF | 67.50 | 80.00 | 55.00 | 0.3615 | 0.3500 | 9 |
| Linear | 72.50 | 80.00 | 65.00 | 0.4551 | 0.4500 | 4-8 |
| POL | 70.00 | 65.00 | 75.00 | 0.4020 | 0.4000 | 18 |
| MLP | 80.00 | 85.00 | 75.00 | 0.6030 | 0.6000 | 20 |
| DNN | 84.50 | 85.00 | 80.00 | 0.6917 | 0.6500 | 12,14 |

Table 3 shows the classification accuracy of vowel /o/ in the same data set for each classifier. Using the DNN classification method, the extracted sound spectrum of the third Mel frequency cepstral coefficient achieves the maximum classification accuracy of 84.5%. Therefore, 33 people are correctly classified, and only 7 are misclassified. Using the same parameters, the maximum sensitivity is 80% and the specificity is 85%. It is concluded that 16 healthy people and 17 PD patients are classified correctly. The maximum MCC is 0.5774 and the PE is 0.7238.

Table 3. Results using vowel /o/

| Classi-fier | Accur-acy(%) | Sensiti-vity(%) | Specifi-city(%) | MCC | PE | Coeffi-cients |
|---|---|---|---|---|---|---|
| RBF | 67.50 | 80.00 | 55.00 | 0.3615 | 0.3500 | 7,8,10 |
| Linear | 72.50 | 80.00 | 65.00 | 0.4551 | 0.4500 | 4,6,7 |
| POL | 67.50 | 70.00 | 65.00 | 0.3504 | 0.3500 | 3 |
| MLP | 77.50 | 80.00 | 75.00 | 0.5507 | 0.5500 | 6 |
| DNN | 84.50 | 80.00 | 85.00 | 0.5774 | 0.7238 | 3 |

Table 4 uses only the vowel /u/ to indicate classification accuracy. Using the DNN classification method, the feature of the extracted sixth Mel frequency cepstrum coefficient is used to achieve the maximum classification accuracy of 89.5%. Therefore, 35 people are correctly classified, and only 5 are misclassified. Using the same parameters, the maximum sensitivity is 80% and the specificity is 95%. From these results, 16 healthy people and 19 PD patients are correctly classified. The maximum MCC is 0.6773 and PE is 0.7440. The example suggests that the vowel /u/ speech sample contains more discriminative information than other speech samples.

Table 4. Results using vowel /u/

| Classi-fier | Accur-acy(%) | Sensiti-vity(%) | Specifi-city(%) | MCC | PE | Coeffi-cients |
|---|---|---|---|---|---|---|
| RBF | 80.00 | 85.00 | 75.00 | 0.6030 | 0.6000 | 9 |
| Linear | 70.00 | 75.00 | 65.00 | 0.4020 | 0.4000 | 4-8 |
| POL | 72.50 | 70.00 | 75.00 | 0.4506 | 0.4500 | 18 |
| MLP | 82.50 | 80.00 | 85.00 | 0.6508 | 0.6500 | 20 |
| DNN | 89.5 | 80.00 | 95.00 | 0.6773 | 0.7440 | 6 |

Table 5 shows the obtained results using all types of voice recordings. These recordings contain the pronunciation of 40 participant vowels /a/, /o/ and /u/, with a total of 120 speech samples(3×40). Using the DNN classification method, the extracted features of the 7-th and 11-th Mel frequency cepstral coefficients achieve the maximum classification accuracy, which is 85.00%. The same parameters give a maximum sensitivity of 85.00% and a specificity of 85.00%, from which it can be concluded that 102 people-time are correctly classified and only 18 are misclassified. It is also shown that 51 healthy people and 51 PD patients are correctly classified. The maximum MCC is 0.6619 and the PE is 0.7080.

Table 5. Results using vowels /a/, /o/ and /u/

| Classi-fier | Accur-acy(%) | Sensiti-vity(%) | Specifi-city(%) | MCC | PE | Coeffi-cients |
|---|---|---|---|---|---|---|
| RBF | 80.00 | 85.00 | 75.00 | 0.6030 | 0.6000 | 19 |
| Linear | 70.00 | 75.00 | 65.00 | 0.4020 | 0.4000 | 16 |
| POL | 72.50 | 70.00 | 75.00 | 0.4506 | 0.4500 | 7 |
| MLP | 82.50 | 80.00 | 85.00 | 0.6508 | 0.6500 | 6 |
| DNN | 85.00 | 85.00 | 85.00 | 0.6619 | 0.7080 | 7,11 |

After training and testing for both groups of subjects (PD patients and healthy individuals), independent data sets containing 28 PD patients are used to test and validate the results. 20 patients with PD and 20 healthy subjects are trained in vowel /a/ and tested from independent data sets. The results are shown in Table 6. The best classification accuracy of SVM and DNN using MLP kernel function and polynomial kernel is 100%. Here, the characteristics of the 5-th, 3-rd, and 7-th Mel frequency cepstral coefficients sequentially achieve this accuracy. This means that 28 patients with PD are classified correctly.

Table 6. Test results using vowel /a/

| Classifier | Accuracy(%) | Coefficients |
|---|---|---|
| RBF | 10.71 | 4,5,6 |
| Linear | 10.71 | 6-10 |
| POL | 100 | 5 |
| MLP | 100 | 3 |
| DNN | 100 | 7 |

The training for vowel /o/ is also performed, and the vowel /o/ provided by the independent data set containing 28 PD patients is tested and verified. It is concluded that the best classification accuracy of the SVM of the polynomial kernel using the characteristics of the 3-rd and 4-th Mel frequency cepstral coefficients and the DNN of the 6-th Mel frequency cepstral coefficient are 100%, as shown in Table 7. This means that 28 patients with PD are successfully classified.

Table 7. Test results using vowel /o/

| Classifier | Accuracy(%) | Coefficients |
|---|---|---|
| RBF | 32.14 | 3 |
| Linear | 89.28 | 3 |
| POL | 100 | 3,4 |
| MLP | 89.28 | 5 |
| DNN | 100 | 6 |

All speech records of the data set are used and experiments are performed at the same time. These recordings contain the pronunciation of the vowels /a/ and /o/ of 28 participants, with a total of 56 speech samples (2×28). After training vowels /a/ and /o/ for 20 PD patients and 20 healthy individuals, the vowels /a/ and /o/ provided by independent datasets of 28 PD patients are tested and verified. As can be seen from Table 8, using the third Mel frequency cepstral coefficient of the DNN has the best classification accuracy of 89.07%. This means that 51 of the 56 PD patients in speech sample are successfully classified, and only 5 people-time are misdiagnosed.

Table 8. Test results using vowels /a/ and /o/

| Classifier | Accuracy(%) | Coefficients |
|---|---|---|
| RBF | 10.71 | 4-8 |
| Linear | 42.85 | 3 |
| POL | 87.50 | 5 |
| MLP | 46.43 | 5 |
| DNN | 89.07 | 3 |

## 5. Discussion

The sound damage of PD patients does not suddenly appear. This is a slow process and the symptoms are at the early stages which may be overlooked. In order to improve the evaluation of Parkinson's disease, the article uses the input speech to calculate the entropy method and take the average value to extract the participants' voiceprint characteristics, and then obtain the parameters of PD patients' dysphonia detection. Compared with the previous RASTA-PLP method, WMFCC has some improvements in comprehensive performance and frame classification. RASTA-PLP characterizes the speech signal by doing short-time spectral analysis. The speech spectrum of RASTA-PLP takes into account the auditory characteristics of the human ear，because the input speech signal is processed by the auditory model, and it facilitates the extraction of anti-noise speech features. However, the cepstrum coefficients extracted by RASTA-PLP contain many frames, which consume lots of processing time in the classification process and hinder accurate diagnosis. HFCC readjusts these cepstral coefficients to quite similar amplitudes by liftering the cepstral coefficients, but the calculation process is complicated and takes a long time. WMFCC has the same anti-noise performance, and can extract more high frequency components from speech signals, so the extracted features have stronger representation ability.

Then the samples in the PD database are trained, and 28 patients with PD are tested and judged. The results show that the characteristics of the sixth Mel frequency cepstral coefficient extracted from the DNN classification method achieves the highest classification accuracy of 89.5%. The accuracy is higher than those with the RBF, Linear, Polynomial and MLP kernel functions of the SVM and PLDA. That is to say the vowel /u/ speech samples contain more discriminative information than other speech samples. The traditional SVM method has better robustness in solving high-dimensional problems, but there is no universal solution to nonlinear problems. The problem lies in the choice of kernel functions and the difficulty in implementing large-scale samples. PLDA is a channel compensation technique base on i-vector. When applied to classification, it is hard to distinguish information in speaker and channel. Compared with the above two classifiers, DNN adopts the layer-by-layer pre-training method based on the restricted Boltzmann machine for the unsupervised pre-training process, and then carries out supervised tuning training. The efficiency of the training is greatly improved, and the problem of local optimum is well improved.When using 28 independent test sets for PD patients, the maximum classification accuracy is always the data from DNN. Studies have shown that DNN can make a better improvement in the classification performance between PD patients and healthy people. It also shows that the DNN network structure is used to enhance the classification ability of voiceprint features.

## 6. Conclusion

WMFCC can effectively extract the high-order cepstral coefficients of the voiceprint, and hence the characterizing ability of the feature components to the audio. Using the MBGD algorithm to optimize the DNN classifier, the computational complexity of the loss function is reduced and the training speed of the system is improved. The experimental results on the PD database show that the classification and recognition models in this paper are superior to the traditional SVM classification methods in terms of sensitivity and accuracy, and provide a new solutions for early diagnosis of Parkinson disease based on voiceprint features.Future work will further study the intrinsic link between the feature extraction of samples and the training of a classifier, classification and complexity of a dataset, and continue to to do some theoretical derivation.

## Acknowledgement

This work is supported by the National Science Foundation of China (Grant No.61673259).

## Conflict of interest statement

We declare that we have no conflicts of interest to this work.We also declare that we do not have any financial and personal relationships with other people or organizations that can inappropriately influence our work.